# Observation of Enhanced Dynamic ΔG effect near Ferromagnetic Resonance Frequency

Wenbin Hu[1], Yudi Wang[1], Mingxian Huang[1], Huaiwu Zhang[1], and Feiming Bai[1]

[1]School of Electronic Science and Engineering, University of Electronic Science and Technology of China, Chengdu 610054, China

* To whom correspondence should be addressed. Electronic mails:

fmbai@uestc.edu.cn

**Abstract**

The field-dependence elastic modulus of magnetostrictive films, also called $\Delta E$ or $\Delta G$ effect, is crucial for ultrasensitive magnetic field sensors based on surface acoustic waves (SAWs). In spite of a lot of successful demonstrations, rare attention was paid to the frequency-dependence of $\Delta E$ or $\Delta G$ effect. In current work, shear horizontal-type SAW delay lines coated with a thin FeCoSiB layer have been studied at various frequencies upon applying magnetic fields. The change of shear modulus of FeCoSiB has been extracted by measuring the field dependent phase shift of SAWs. It is found that the $\Delta G$ effect is significantly enhanced at high-order harmonic frequencies close to the ferromagnetic resonance frequency, increasing by ~82% compared to that at the first SAW mode (128 MHz). In addition, the smaller the effective damping factor of magnetostrictive layer, the more pronounced $\Delta G$ effect can be obtained, which is explained by our proposed dynamic magnetoelastic coupling model.

The $\Delta E$ or $\Delta G$ effect is known as the modification of Young's modulus or shear modulus of a magnetostrictive (MS) material with respect to a magnetic field. Although giant $\Delta E$ or $\Delta G$ effect exists in bulk amorphous magnetostrictive materials with very low anisotropy, their counterparts in a thin-film form are significantly reduced due to structural inhomogeneity, pinning defects, and surface or interface stress. For example, the maximum $\Delta E$ effect of FeSiB Metglass ribbons is 64%, and its magnetoelastic modulus coefficient (MMC) ($dE/dH$) reaches 150 GPa/Oe[1]. But the maximum $\Delta E$ effect and the $dE/dH$ of FeCoSiB films are only 30% and 6.5 GPa/Oe, even after systemically process optimization[2].

Recently, a lot of magnetic field sensors have been developed utilizing either cantilever beams[3–5], or surface acoustic wave (SAW) resonators[6–12] and delay lines[13–16]. Taking magnetic SAWs for example, the phase velocity of the piezoelectric substrate becomes dispersive upon coating a magnetostrictive layer onto or in-between interdigital transducer electrodes (IDTs). Thus, magnetic fields can be detected by tracing the frequency or phase shift of SAW resonators or delay lines. Rayleigh-type SAWs heavily rely on the magnitude of $\Delta E$ effect, while shear horizontal (SH)-type SAWs demand a large $\Delta G$ effect.

In early studies of the $\Delta E$ or $\Delta G$ effect[17–19], a static magnetization assumption was applied, i.e. magnetic domains/moments can immediately rotate to their equilibrium orientation upon applying an external field or stress. However, this is not true when the SAW frequency approaches the ferromagnetic resonance (FMR) or spin wave resonance (SWR) frequency, which is typically hundreds of MHz or even GHz for soft magnetic film. There are few studies on frequency-dependence of $\Delta E$ or $\Delta G$. H. Zhou et al.[13] studied the phase velocity shift of the 1st and 3rd order Rayleigh waves and SH waves in [TbCo2/FeCo]$_{20}$/Y-cut

LiNbO$_3$ multilayers, and found that the velocity shift of the SH wave increased from 0.2% (270 MHz) to 0.6% (822 MHz). Similarly, A. Mazzamurro et al.[15] used a [TbCo$_2$/FeCo]$_{25}$/ST-X90°-cut quartz structure to obtain the phase velocity shifts of Love-type SAWs under the 1st (410 MHz) and 3rd (1.2 GHz) excitations. M. Elhosni et al.[20] investigated the center frequency shift in a Ni/ZnO/IDTs/LiNbO$_3$ structure under excitation of the 1st, 3rd, and 5th order Rayleigh waves. Spetzler et al.[21] derived the theoretical frequency-dependence of $\Delta E$ effect and concluded that the $\Delta E$ effect reduces with increasing frequency. However, these works either focused on SAW frequency far lower than FMR frequency, due to the very high anisotropic field of the TbCo$_2$/FeCo film, or used magnetostrictive films (like Ni) with a large damping factor.

In this work, we report the frequency-dependence of $\Delta G$ effect in amorphous Fe-Co-Si-B film with a large saturation magnetostriction coefficient up to 55 ppm [22] and a low Gilbert damping factor down to 0.0038 [23]. A split-finger IDTs design was employed to excite high-order harmonic SAWs, which is sufficient to cover a wide frequency band from 128 MHz to 1.9 GHz. The field- and frequency-dependence of shear modulus was successfully extracted by monitoring the phase shift of SH-type SAWs and separating the geometry and structure contributions through finite element analysis.

Fig. 1(a) schematically illustrates our SAW delay line on a 42°-rotated Y-cut LiTaO$_3$ (TDG, Zhejiang, China) substrate, which supports SH-type SAWs[24]. Aluminum split-finger interdigital IDTs with a thickness of 50 nm were designed to reduce the reflection of SAWs and excite higher order harmonics. The periodicity is 32 μm, and the distance between the two IDTs is 1.2 mm, as labeled in the figure. A 1 mm × 1.2 mm rectangular

$(Fe_{90}Co_{10})_{78}Si_{12}B_{10}$ thin film was then sputtered between them. Two different thicknesses (30 and 75 nm) were deposited, and the corresponding device are labeled as Device 1 and Device 2, respectively. During the deposition of Fe-Co-Si-B films, an in-situ magnetic field was applied parallel to the SAW propagation direction ($\vec{k}$) to induce in-plane uniaxial anisotropy. Fig. 1(b) shows the optical image of Device 1. The two IDTs were then connected to 50-Ω microstrip lines on a PCB board via silver wires. Both ends of the microstrip lines were soldered to SMA connectors, as shown in Fig. 1(c). The transmission parameters $S_{21}$ of the delay lines were measured by a vector network analyzer (VNA, Agilent N5230A). The device was placed in a Helmholtz coil controlled by a DC current source (ITECH-6502), the output magnetic field was calibrated using a Gaussmeter (Lake Shore 425). All tests were automatically controlled by a LABVIEW program.

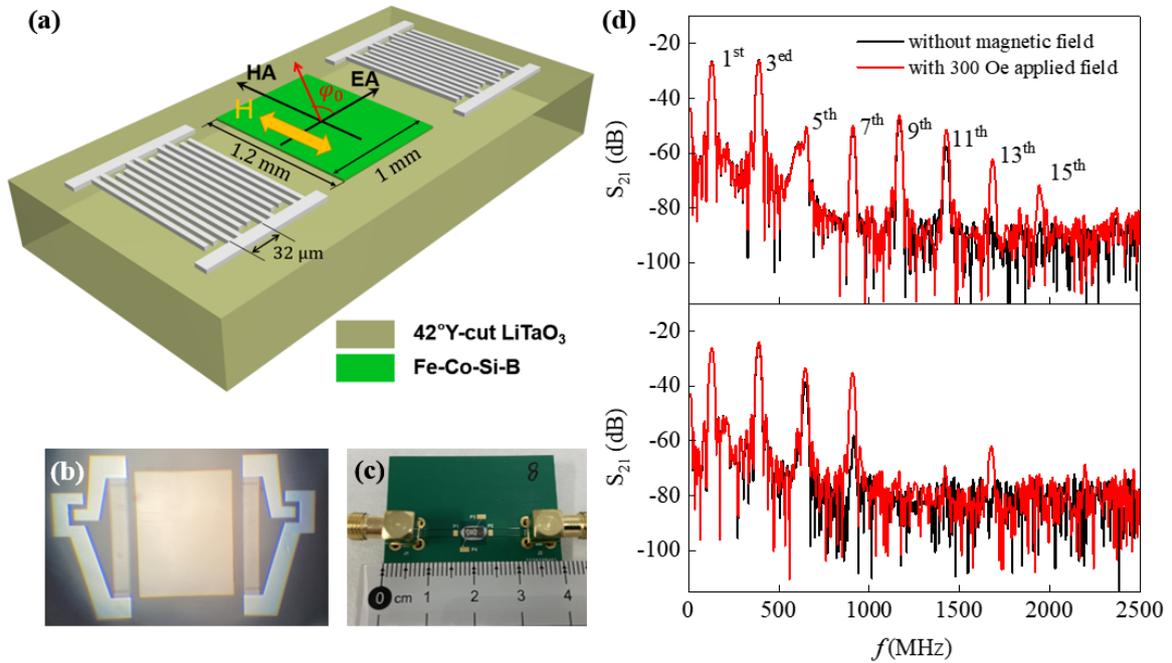

Fig. 1 (a) Schematic illustration of a SAW delay line on a 42°-rotated Y-cut $LiTaO_3$ substrate. (b) and (c) Photographs of the fabricated and packaged SAW device. (d) Measured $S_{21}$ for Device 1 (top) and Device 2 (bottom) with and with no applied field of 300 Oe.

Firstly, a wideband measurement from 10 MHz to 3 GHz was performed, and the $S_{21}$ magnitude of Device 1 and Device 2 with and with no saturated magnetic field ($H_s = 300$ Oe) are compared and shown in Fig. 1(d). Note that each measured $S_{21}$ has been subjected to a time-domain gating[25] process to isolate the electromagnetic crosstalk between IDTs. In the absence of external magnetic fields, Device 1 shows the first mode and extra five odd harmonics, with the center frequency $f_{SAW} = qf_0$, where $f_0 = 128$ MHz, and $q(= 1, 3, …, 11)$ represents the corresponding harmonic modes. The insertion loss of each harmonic decreases when $H_s$ is applied, especially for the higher harmonics, thus, additional 13$^{th}$ and 15$^{th}$ harmonics appear. Device 2 shows four harmonic modes with $f_0 = 128$ MHz and $q = 1, 3, 5, 7$ under zero magnetic field, and the insertion loss also decreases significantly upon applying a 300 Oe field.

The phase of SAWs can be extracted at the center frequency of each mode using the formula below

$$\phi = arctan\left(\frac{R(S_{21})}{I(S_{21})}\right). \qquad (1)$$

This allows us to carry out a detailed study about the field-dependence of SAWs phase shift at the 1$^{st}$, 3$^{rd}$, ..., and 11$^{th}$ harmonics. A cycling magnetic field was applied along the hard axis (perpendicular to $\vec{k}$) of the Fe-Co-Si-B film, starting from 0 to 80 Oe, followed by a decrease to -80 Oe, and then returning to 80 Oe. The relative phase shift is defined by subtracting the phase under a saturated field as $\Delta\phi(H) = \phi(H) - \phi(H_s)$. The results for all modes of Device 1 are shown in Fig. 2. An approximately 1.3 Oe deviation is observed between the measured $\Delta\phi$-$H$ curves during increasing and decreasing magnetic fields, which can be attributed to the small coercivity of Fe-Co-Si-B film. The measured $\Delta\phi(H)$ under zero field is only about -0.8° for the first mode, but dramatically increases up to -203° for the 11$^{th}$ harmonic. We have also plotted the phase sensitivity $d\phi/dH$ on the right axis of Fig. 2. There are clearly four $d\phi/dH$ peaks for the first mode, two locating at a small bias field $H_b$ of $\pm$13.8 Oe and the other two at a relatively large $H_b$ of $\pm$30.2 Oe. With increasing harmonic frequency, both values increase, but the former become more distinct. For the 11$^{th}$ harmonics, the maximum phase sensitivity reaches 19.6°/Oe or 196°/mT. Therefore, we will focus on the phase sensitivity at the smaller bias field in the followed discussion.

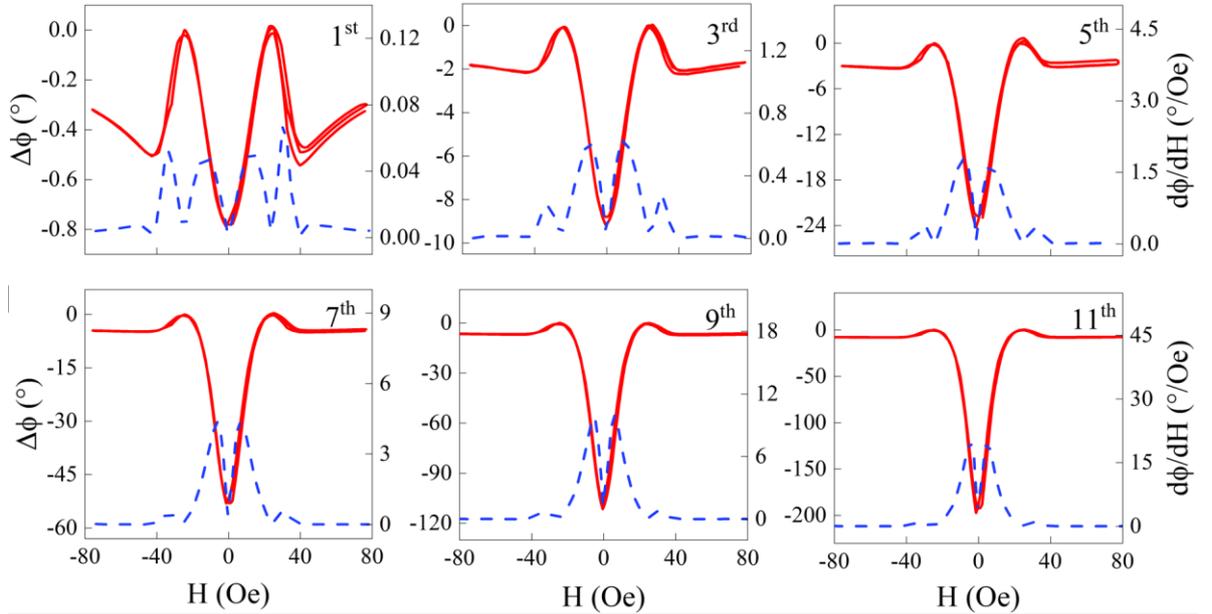

Fig. 2 Measured field dependent phase shift and phase sensitivity of Device 1 from the 1$^{st}$ to the 11$^{th}$ order mode.

The field dependent phase shift of Device 2 shows a similar trend like Device 1, with -4.2° and -355.4° phase shift, respectively, for the first mode and the 7$^{th}$ harmonic (not shown here). Fig. 3 plots the maximum phase sensitivity $(d\phi/dH)_{max}$ and its corresponding bias field $H_b$ for Device 1 and Device 2 as a function of SAW frequency. Similar with $\Delta\phi$, the magnitude of $(d\phi/dH)_{max}$ increases exponentially with frequency, partially due to the shorter wavelength of higher harmonic SAWs. Notice that the thickness of MS layer plays an important role in deciding both $(d\phi/dH)_{max}$ and the corresponding $H_b$. As seen in Fig. 3, Device 2 exhibits much higher $(d\phi/dH)_{max}$ due to the thicker MS layer (75 nm) employed, which reaches 134.5°/Oe or 1345°/mT for the 7[th] harmonic. Meanwhile, the $H_b$ of Device 1 is always larger than that of Device 2. This can be attributed to the larger surface anisotropy in the thin ferromagnetic film (30 nm)[26,27]. Interestingly, a clear decreasing trend of $H_b$ was observed in both devices, with Device 1 decreasing from 13.8 to 4.4 Oe and Device 2 decreasing from 3.0 to 1.4 Oe. This low bias field is desired for weak magnetic field sensors and magnetic field tuned delay lines.

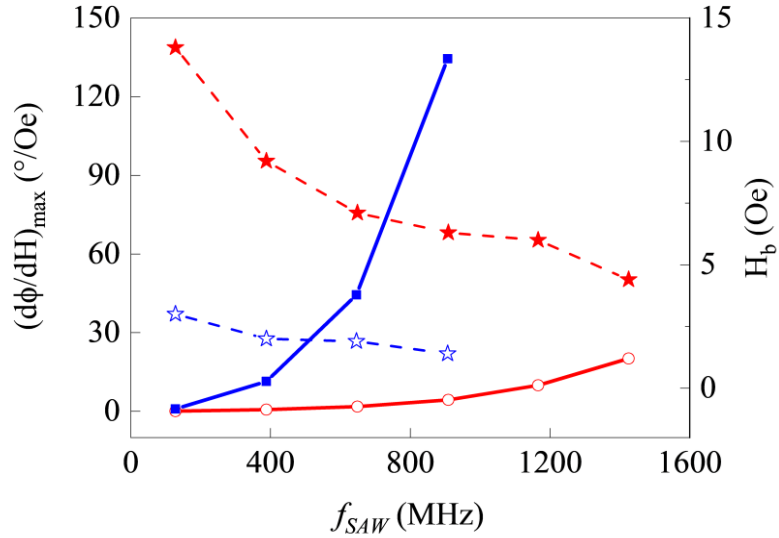

Fig. 3 Measured frequency-dependence of the maximum phase sensitivity (solid line) and the corresponding bias field (dashed line) for Device 1 (red) and Device 2 (blue).

As mentioned earlier, the phase shift of SH-type SAWs arises from the $\Delta G$ effect of Fe-Co-Si-B film, which in turn affects the phase velocity of SAWs and causes the $S_{21}$ phase shift. But measuring $G$ through the SAWs phase shift is complicated due to the influences of the multilayer structure, the geometries of IDTs and MS layer, and the wavelength of SAWs[14]. To extract the field-dependence of $G$, it is necessary to identify and separate the influence of these factors.

The phase $\phi(H)$ of SAWs is a function of phase velocity $v(H)$ and the length ($L$) of Fe-Co-Si-B film, and can be expressed as

$$\phi(H) = \frac{2\pi f_{SAW}}{v(H)} L, \qquad (2)$$

So, $\Delta\phi(H)$ in Fig. 3 can be converted to the phase velocity using formula

$$v(H) = \frac{2\pi f_{SAW} L}{2\pi f_{SAW} L/v(H_s) + \Delta\phi(H)}, \qquad (3)$$

with $v(H_s)$ as the SAW phase velocity under a saturation field. As can be seen, it is necessary to know $v(H_s)$ before determining $v(H)$.

In our case, when SAWs propagate into Fe-Co-Si-B/LiTaO$_3$ multilayered, the phase velocity becomes dispersive, depending on the ratio between the thickness $d$ of Fe-Co-Si-B and the SAW wavelength $\lambda$. Therefore, $v(H_s)$ has different values for each mode of Devices 1 and 2. A finite element model (FEM) was then built to simulate $v(H_s)$ in a Fe-Co-Si-B/LiTaO$_3$ multilayered structure with various values of $d/\lambda$, as shown in Fig. 4. For magnetic field saturated Fe-Co-Si-B, we used a Young's modulus of 150 GPa, a Poisson's ratio of 0.3 and a volume density of 7250 kg/m$^3$, which results in $G(H_s) = 57.7$ GPa and SH-wave phase velocity $v_{MS} = 2821$ m/s. The elastic parameters of LiTaO$_3$ substrate are available in the COMSOL Multiphysics software, and the calculated SH-wave phase velocity ($v_{LT}$) is 4080 m/s. According to the pioneer work of Farnell and Adler[28], the phase velocity of multilayered SAW lies between $v_{LT}$ and $v_{MS}$. As $d/\lambda$ increases, $v(H_s)$ gradually deviates from $v_{LT}$ and approaches $v_{MS}$. The zoom-in insert in Fig. 4 also highlights the $v(H_s)$ values corresponding to all harmonic modes of Devices 1 and 2.

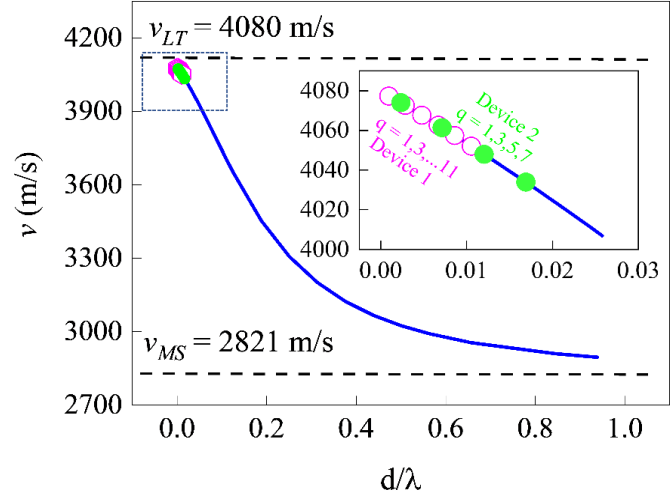

Fig. 4 Simulated phase velocity as a function of $d/\lambda$ for SH-SAWs propagating in the Fe-Co-Si-B/LiTaO$_3$ multilayered structure. Insert shows $v(H_s)$ of each mode present in Devices 1 and 2.

Now, for a fixed $v(H_s)$ of each mode, the variation of the shear modulus $G$ of Fe-Co-Si-B with applied fields can be determined via inputting a series of $G$ values into the FEM until the simulated phase velocity matches that obtained by using Eq. (3). The extracted $G$ versus $H$ results are plotted in Fig. 5(a) and 5(b). Here, we express the $\Delta G$ effect as $(G - G(H_s))/G(H_s) \times 100\%$. Our results indicate that both Device 1 and Device 2 exhibit a frequency dependent $\Delta G$ effect, as demonstrated by the fact that the $\Delta G$ effect of Device 1 at zero field increases from 11% to 20%, while the counter part of Device 2 exhibits an increase from 23% to 39%.

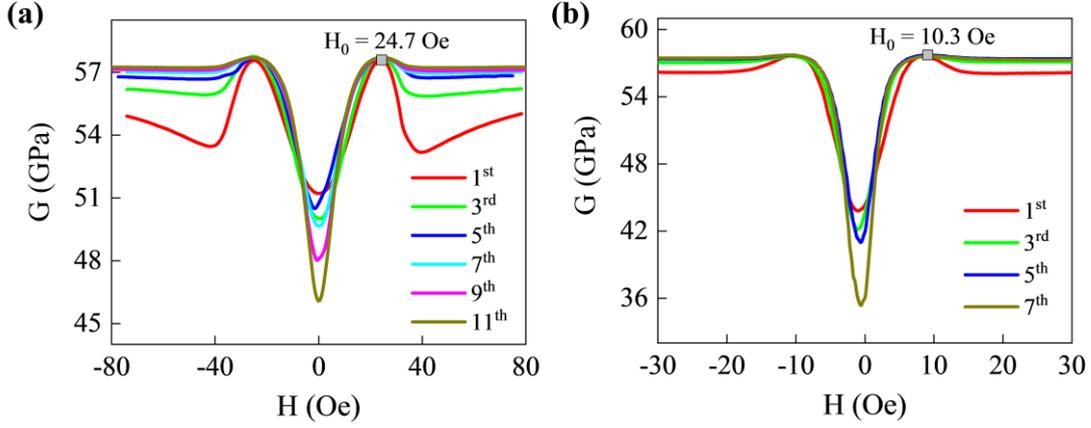

Fig. 5 *G-H* curves for various modes in (a) Device 1 and (b) Device 2, respectively.

This distinct feature can be explained by a dynamic magnetoelastic coupling model recently developed by our group[11]. The shear modulus $G$ can be written as a function of the external magnetic field $H$ and the SAW frequency $f_{SAW}$

$$G(H, f_{SAW}) = G(H_s) + \frac{B_2^2 \cos^2 2\varphi_0}{2\mu_0 M_s} \frac{\gamma(i\omega\alpha + \omega_m)}{\omega^2 - \left(i\omega\alpha + \gamma H \sin\varphi_0 + \gamma H_k \cos 2\varphi_0 + \omega_m \frac{kd}{2}\sin^2\varphi_0\right)(i\omega\alpha + \omega_m)} \quad (4)$$

with

$$\omega = 2\pi f_{SAW}, \omega_m = \gamma M_s, \sin\varphi_0 = \begin{cases} H/H_k, & |H| \leq H_k \\ 1, & |H| > H_k \end{cases}.$$

where $M_s$, $\gamma$, $H_k$, $B_2$, and $\alpha$ are the saturation magnetization, gyromagnetic factor, uniaxial anisotropy field, magnetoealstic coupling coefficient, and damping factor. $\varphi_0$ is the angle between the equilibrium magnetization and the SAW propagation direction, as shown in Fig. 1(a). Note that all *G-H* curves in Fig. 5 reach their peaks at the same applied field $H_0$. Since $G(H, f_{SAW}) - G(H_s)$ is proportional to $\cos^2 2\varphi_0$, $G(H_0, f_{SAW})$ is equal to $G(H_s)$ when $\varphi_0 = 45^o$, therefore, the uniaxial anisotropy field $H_k$ can be calculated as $H_k = \sqrt{2}H_0$.

Eq. (4) can be also used to determine the magnetoelastic coupling coefficient $B_2$, which is related to the stress state, and the damping parameter $\alpha$, which is sensitive on microstructure, by setting $H = 0$ Oe, and rewritten as

$$G(0\,\text{Oe}, f_{SAW}) = G(H_s) + \frac{B_2^2}{2\mu_0 M_s} \frac{\gamma(i\omega\alpha + \omega_m)}{\omega^2 - (i\omega\alpha + \gamma H_k)(i\omega\alpha + \omega_m)}. \quad (5)$$

When $f_{SAW}$ is far lower than the FMR frequency, $f_{FMR} = \gamma\sqrt{H_k M_s}/2\pi$ (2.1 Ghz for Device 1, and 1.4 GHz for Device 2 using parameters in Table I), Eq. (5) can be further simplified as

$$G(0\,\text{Oe}, f_{SAW}) = G(H_s) - \frac{B_2^2}{2\mu_0 M_s H_k}, \text{for } f_{SAW} \ll f_{FMR}. \quad (6)$$

Apparently, $B_2$ can be calculated as $\sqrt{2\mu_0 M_s H_k [G(H_s) - G(0\,\text{Oe}, f_{SAW})]}$. In our previous work [23], we have already determined the saturation magnetization ($M_s = 15.3$ kOe) and the gyromagnetic factor ($\gamma = 18.3$ MHz/Oe) of Fe-Co-Si-B film. Comparing Device 1 with Device 2, one can find that the $\Delta G$ effect is inversely proportional to $H_k$, as expect from Eq. (6).

Moreover, we can fit α using a series of $G(0\,\text{Oe}, f_{SAW})$ at their harmonic frequencies via Eq. (5). The frequency-dependence of $G(0\,\text{Oe}, f_{SAW})$ of both Device 1 and Device 2 are plotted in Fig. 6 using different α values. One can observe a sharp change in the $\Delta G$ effect near $f_{FMR}$, particularly when α is set to a low value. Now, it turns clear that the damping factor plays a critical role in determining the magnitude of the ΔG effect. A low damping factor corresponds to an enhanced $\Delta G$ effect when $f_{SAW}$ is close to $f_{FMR}$, whereas a large damping factor leads to the reduction of $\Delta G$ effect. The fitted values of α are 0.008 for Device 1 and 0.01 for Device 2, respectively. The $\Delta G$ effect of Device 2 at its 7[th] harmonic mode (896 MHz) shows an increase of 70% compared to that at the first mode (128 MHz). An even higher enhancement of 82% is observed for Device 1 due to its lower damping factor. All these fitted results for Devices 1 and 2 are listed in Table I.

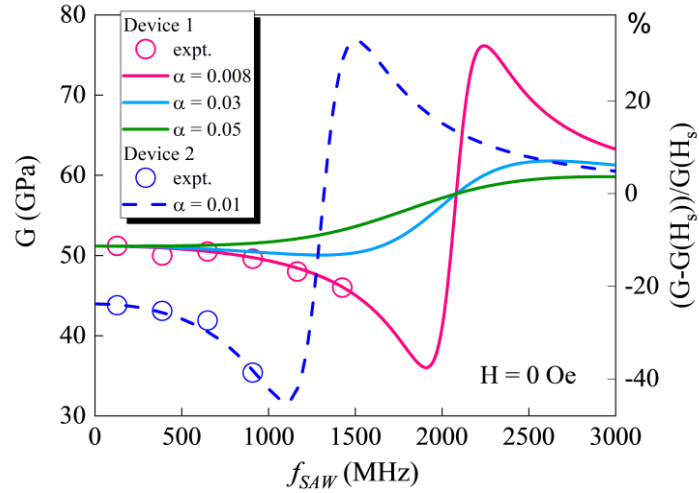

Fig. 6 Frequency-dependence of shear modulus of Fe-Co-Si-B film of Device 1 and Device 2 under zero magnetic field. Here, lines represent the calculated results via Eq. (5) with different damping factor settings, and dots correspond to experimental results.

Table I. Summary of the parameters for Devices 1 and 2 obtained by fitting the *G-H* curves.

| Device | $M_s$ (kOe) | $\gamma$ (MHz/Oe) | $H_k$ (Oe) | $B_2$ (MPa) | $\alpha$ |
|--------|-------------|-------------------|------------|-------------|----------|
| 1      | 15.3        | 18.3              | 35         | 7.5         | 0.008    |
| 2      | 15.3        | 18.3              | 14.5       | 7.0         | 0.01     |

In summary, the frequency-dependence of shear modulus of Fe-Co-Si-B thin films has been studied by measuring the phase shift of SH-type SAW delay lines. It is found that for the same device structure and MS layer dimensions, high-order harmonic SAW excitations can significantly enhance the $\Delta G$ effect through magnetoacoustic coupling. Moreover, a lower damping factor corresponds a stronger $\Delta G$ effect, which can be explained based on a dynamic magnetoelastic model. Our findings not only manifest a novel dynamic $\Delta G$ effect, but also offer a very useful method to boost the sensitivity of magnetic field sensors based on surface acoustic waves or bulk acoustic waves.

**Acknowledgement**

This work is supported by the National Natural Science Foundation of China (Grant No. 61871081) and the Natural Science Foundation of Sichuan Province under Grant No. 2022NSFSC0040.